# Variations of CR-muon intensity in the declining phase of the 23$^{rd}$ solar cycle in ground and shallow underground data

A. Dragić[a], R. Banjanac[a], V. Udovičić[a], D. Joković[a], J. Puzović[b] and I. Aničin[b]
*(a) Institute of Physics, Pregrevica 118, Belgrade, Serbia and Montenegro*
*(b) Faculty of Physics, University of Belgrade, Studentski trg 12-16, Belgrade, Serbia and Montenegro*
Presenter: R. Banjanac (radomir.banjanac@phy.bg.ac.yu), yug-banjanac-R-abs1-sh15-poster

The cosmic-ray intensity record from Belgrade muon detectors in the period 2002-2004 is subjected to power spectra analysis. Statistically significant peaks are found in datasets from both ground and underground (25 m w.e.) detectors. The possible origin of these periodicities is discussed.

## 1. Introduction

The cosmic-rays (CR) arriving at the Earth after propagation through the heliosphere carry information on the interplanetary magnetic field. The structure of that field, on the other hand, varies under the influence of phenomena in which solar activity is manifested. Therefore, variation of CR flux is expected to be good indicator of solar activity.

The CR time series have been analyzed in a search for periodic intensity variations by various authors. Majority of studies is with neutron monitor records, covering lower CR energy than muon detectors. Generally, a few studies are dedicated to spectral analysis of muon time series. Muons are produced by higher energy primaries than neutrons, and therefore muon flux variability is a valuable tool for better understanding of solar modulation and connection between solar activity and cosmic-rays. Of course, cosmic-ray and solar variability can not be related in a straightforward manner, due to complexity of mechanism of solar modulation of cosmic rays.

Among numerous studies of solar variability periodicities which are reported, there are identified periodicities in coronal mass ejection data from LASCO/SOHO during maximum of solar cycle 23 (1999-2003) [1]. This result is highly relevant for the present study, since time interval investigated partially overlaps with ours and since relation between CME and cosmic rays is well established [2].

The new results of our continuous measurement of cosmic-ray muons that has started in 2002, [3], for the period 2002-2004 yield statistically significant improvements in the interpretation of observed periodicities.

## 2. Experiment Description

Cosmic-ray muons are detected by two identical plastic scintillator detectors, installed in the Institute of Physics, Zemun, Belgrade, Serbia & Montenegro. The laboratory is located at geographic latitude 44$^{o}$51'N; geographic longitude 20$^{o}$23'E, and altitude 78m a.s.l. Geomagnetic latitude of the laboratory is 39$^{o}$32'N and vertical geomagnetic rigidity cut-off is 5.3GV.

One detector is situated on the ground level, while the other is in the "Dr. Radovan Antanasijević" underground laboratory (at the depth of 25m water equivalent). The detectors are of prismatic shape with dimensions 50cm x 23cm x 5cm. Detectors are produced by the High Energy Physics Laboratory of JINR, Dubna, and scintillator type is similar to NE102. Each detector lies horizontally on its largest side and single 5 cm photomultiplier watches its long side (50cm x 5cm) via a correspondingly shaped light guide.

After amplification the analog output signal is digitized by laboratory made A/D converter and then linked to a computer PCI card. The 4096 channel spectrum is automatically recorded every 5 min, with 270 sec dedicated to measurements, and 30 sec being allowed for recording on local hard disc, some quick interventions on the system, and data transmission to second local network computer. The setup enables off-line data analysis without interrupting the measurements.

The recorded spectrum is mainly the spectrum of muon energy deposit ΔE. Daily recorded spectrum from surface detector, together with Monte Carlo simulation of detector response is plotted in Figure1. The spectrum stretches to about 200 MeV and has a well defined single particle peak corresponding to an energy loss of some 11 MeV. A peak in the low energy part of the spectrum is due to ambient radiation.

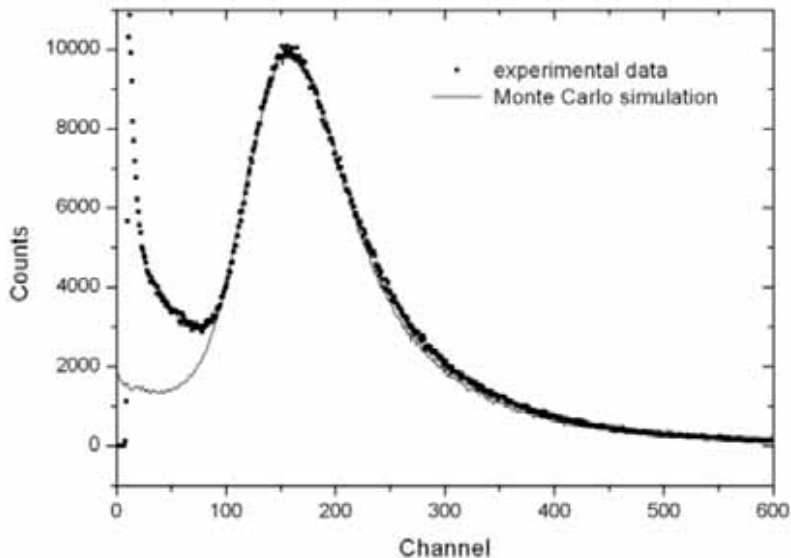

**Figure 1.** Energy loss spectrum in plastic scintillator detector and Monte Carlo simulation of detector response to CR muons. The low energy peak is due to ambient radiation, whose contribution to overall spectrum is well separated from that of muons.

Monte Carlo simulation of this ΔE spectrum, based on GEANT4 package, [4], agrees with the experimental spectrum within statistical error. With the given geometry, detector response to cosmic and ambient radiation is well separated (as confirmed by MC simulation) and unambiguous YES/NO selection criteria can be applied. In the simulation, $\cos^2\theta$ directional distribution of incoming muons is assumed. All relevant physical processes leading to muon energy loss in interaction with detector is taken into account (ionization, bremsstrahlung, and pair creation). Contribution to total energy loss from δ-electrons and other particles, produced by muonic interaction is also accounted for. Multi-muon events, resulting from extensive air showers are included in simulation as well. Reliable detector simulation allows improvement of data analysis.

## 3. Data analysis and main results

The data from Belgrade muon detectors for the period 2002-2004 (descending phase of 23$^{rd}$ cycle) are spectrally analyzed. The time interval covers years following maximum of solar cycle 23, when the Sun was exceptionally active.

Lomb-Scargle periodogram, [5], analysis method has been used in spectral analysis of cosmic muon time series. This particular method is preferred since it can successfully treat unevenly sampled and gapped time series. Another advantage of Lomb-Scargle method is well known statistical interpretation of periodogram.

Better definition of peak position and more smooth appearance of periodogram is achieved by oversampling (the number of sampled frequencies in periodogram is larger than the number of independent frequencies). Nevertheless, the frequency resolution is determined by the spacing between independent frequencies. In our case, with 4 hour averaged data, the investigated frequency range is between Nyquist critical frequency (34μHz) and inverse of total time span (10.6nHz). The number of independent frequencies is N=3427 and frequency resolution is δν=10nHz. If the individual peak width is larger than frequency resolution it is a signature of a quasi-periodicity with variable period, rather than a true periodicity.

Beside statistical, periodogram analysis faces also spectral problems (spectral leakage and aliasing). Thus, statistical criterion alone, based on false alarm probability is not sufficient for discrimination between true and false periodicities in the time series. One way to address this problem is by harmonic filtering, [6]. After calculating periodogram, usually strongest signal is subtracted from the data to test its influence on other signals and then periodogram is recalculated.

Yet another problem might plague the result - the pattern in missing data could cause the presence of false peaks and in the case of noisy data could shift position of true peaks.

Above mentioned problems are addressed by CLEAN deconvolution algorithm, [7]. The true, undistorted spectrum is obtained by deconvolution of "dirty" spectrum from spectral window function. The spectrum referred to as "dirty" is computed in our case as Schuster periodogram and it turns out to be almost identical to Lomb-Scargle periodogram. Deconvolution procedure is started from highest amplitude component in "dirty" spectrum. A fraction of its amplitude (named gain: 0<g<1) is convoluted and removed from "dirty" spectrum, resulting also in removal of its sidelobs in residual spectrum. In our calculations value g=0.1 is used, but results are not sensitive to change of this value. Residual spectrum is processed in the same manner until the stopping condition is met. We have chosen stopping condition recommended by [8], that residual spectrum is not significantly different from pure noise.

The statistically significant peaks identified in the spectra are listed in Table 1 and Table 2. The $\Delta T_a$ is an error estimated theoretically, knowing frequency resolution (δν=10nHz), while $\Delta T_b$ is experimentally determined as a half of full width at half maximum of a given signal. These errors are comparable, making it difficult to distinguish between true and quasi-periodicities.

Table 1. Periodicities present in the underground data set. Periods are given in days.

| T | 1 | 8.7 | 13.6 | 20.5 | 26.5 | 34.5 | 37 | 77 | 90 | 162 | 194 | 236 | 350 |
|---|---|-----|------|------|------|------|----|----|----|----|-----|-----|-----|-----|
| $\Delta T_a$ | $5 \cdot 10^{-4}$ | 0.03 | 0.1 | 0.2 | 0.3 | 1 | 0.6 | 2.5 | 3.5 | 11.5 | 16 | 24 | 53 |
| $\Delta T_b$ | - | 0.1 | 0.1 | 0.2 | 0.3 | 0.5 | 0.5 | - | 4 | 10 | 20 | 23 | 45 |

Table 2. Periodicities present in the ground data set.

| T | 5.3 | 8.4 | 13.6 | 20.5 | 27 | 34.6 | 37 | 57 | 90 | 194 | 237 | 350 |
|---|-----|-----|------|------|----|------|----|----|----|-----|-----|-----|
| $\Delta T_a$ | $5 \cdot 10^{-4}$ | 0.03 | 0.1 | 0.2 | 0.3 | 1 | 0.6 | 2.5 | 3.5 | 16 | 24 | 53 |
| $\Delta T_b$ | - | 0.1 | 0.1 | 0.2 | 0.5 | 0.5 | 0.6 | - | 4 | 15 | 18 | 43 |

A number of statistically significant periodicities are identified by spectral analysis. Most of them are common features of both data sets, but several are unique to one or another detector data. These peaks are particularly interesting, since they appear as a consequence of energy dependent modulation processes. Namely, muons detected at surface have energy threshold of ~1GeV, while underground detector is reached by muons having energy higher than 5GeV at ground level, and are produced by primaries of higher energy.

A periodicity with solar rotation period (~26.6 ± 0.3 day in underground and ~27.0 ± 0.3 day in ground data) is easily identified. This signal is well documented in various parameters describing solar activity and it is also present in many cosmic-ray time series. By far less significant are higher harmonics of solar rotation period (~13.6 ± 0.1 in both data sets and ~8.70 ± 0.03(~ 8.40 ± 0.03) days). The presence (and absence) of these signals might be related to sectorial structure of the interplanetary magnetic field.

In the high frequency region there is a peak at 1day period in the underground data. The same signal is recognized by Lomb-Scargle periodogram, slightly above 99% confidence level, but the signal is not confirmed by CLEAN. This periodicity is partly due to atmospheric effects but also indicates CR anisotropy arising from co-rotation of CR particles with interplanetary magnetic field lines. In the surface detector data some evidence is found for ~3.5d and ~5.3d signals, both being higher harmonics of solar rotation period.

A set of periodicities is present in the region of intermediate frequencies: from 20.5 ± 0.2d to 90 ± 3.5d. Interestingly, the signal with solar rotation period does not appear to be the most pronounced. Proliferation of peaks in this frequency band came as a surprise and one might question whether they represent genuine signal or artifacts of methods of analysis. CLEAN indeed removed several peaks, significant at 99% level in Lomb-Scargle periodogram but the above mentioned remained. The only known case where CLEAN fails is when false signal has the highest amplitude in periodogram.

In the low frequency region, 162 ± 10 day periodicity is found only in UD. This period correspond to six solar rotations (6 x 27 = 162 day). It is not clear whether this signal can be correlated with variations of many solar parameters with periods ranging from 154 to 158 day but there are some predictions that this periodicity relates to the strong magnetic field via influence of magnetic clouds on CR intensity variation.

Similar periodicities to those we found are reported in various solar activity parameters, confirming the value of cosmic-ray studies as an indirect diagnostic tool of solar activity and represent a strong motive for further investigation of relation between solar activity and CR variations.

## 4. Acknowledgements

Generous support from Ministry of Science and Environmental Protection of Republic of Serbia (project No. 1461) is gratefully acknowledged. The Belgrade cosmic-ray group is especially grateful to Sir Arnold Wolfendale for his constant interest and help in this work.

## References

[1] Y.Q. Lou et al., Mon. Not. R. Astron. Soc. 345, 809 (2003).
[2] H. V. Cane, Space Science Reviews. 93, 55 (2000).
[3] J. Puzović et al., 28th ICRC, Tsukuba (2003) 1199.
[4] S. Agostinelli et al., Nucl. Instr. & Meth. A 506, 250 (2003).
[5] N.R.Lomb, Ap&SS 39, 447 (1976), J.D.Scargle, ApJ 263, 835 (1982).
[6] S.Ferraz-Mello, AJ 86, 619 (1981).
[7] D.H.Roberts et al., AJ 93, 968 (1987).
[8] V.V.Vityazev, Analiz neravnomernih vremennih ryadov (in Russian), Sankt-Peterburg (2001).